\documentstyle[12pt,epsf]{article}
 
\def\p{\par\noindent\phantom b}
\newcommand{\eq}{\begin{equation}}
\newcommand{\en}{\end{equation}}

\newcommand{\eqa}{\begin{eqnarray}}
\newcommand{\ena}{\end{eqnarray}}
\newcommand{\eqan}{\begin{eqnarray*}}
\newcommand{\enan}{\end{eqnarray*}}

\newcommand{\lbl}{\label}

\newcommand{\sect}[1]{\setcounter{equation}{0}\section{#1}}

\newcommand{\AP}[1]{Ann. Phys.\ {\bf #1}\ }

\newcommand{\JMaP}[1]{J. Math. Phys.\ {\bf #1}\ }

\newcommand{\NP}[1]{Nucl. Phys.\ {\bf #1}\ }

\newcommand{\PL}[1]{Phys. Lett.\ {\bf #1}\ }
\newcommand{\PR}[1]{Phys. Rev\ {\bf #1}\ }

\newcommand{\PRL}[1]{Phys. Rev. Lett.\ {\bf #1}\ }
\newcommand{\RNC}[1]{Riv. Nuovo Cimento\ {\bf #1}\ }

\def\sqr#1#2{{\vcenter{\hrule height.#2pt
     \hbox{\vrule width.#2pt height#1pt \kern#1pt
        \vrule width.#2pt}
     \hrule height.#2pt}}}

\def\thinspace{\kern .16667em}

\def\Dir{\nabla\kern-2ex\Big{/}}

\def\Dsl{\partial\kern-1.5ex\Big{/}}
\def\psl{{p\kern-1ex {/}}}
\def\qsl{{q\kern-1ex {/}}}
\def\ksl{{k\kern-1.2ex {/}}}

\def\llog{\log\left({\Lambda^2\over m^2}\right)}
\def\atan{{\rm atan}}
\def\Li{{\rm Li}}

\def\api{{\alpha\over\pi}}
\def\za{Z\alpha}
\def\sq2{\sqrt{2}}

\def\reali{{\hbox{\s@ l\kern-.5ex R}}}
\def\naturali{{\hbox{\s@ l\kern-.5ex N}}}
\def\interi{{\mathchoice
 {\hbox{Z\kern-1.5mm Z}}
 {\hbox{Z\kern-1.5mm Z}}
 {\hbox{{Z\kern-1.2mm Z}}}
 {\hbox{{Z\kern-1.2mm Z}}}  }}

\def\unity{{\hbox{\s@ 1\kern-.8mm l}}}
\def\uno{{\hbox{ 1\kern-.8mm l}}}
\def\pd#1{{\partial~\over\partial #1}}

\def\part{\partial}

\def\aa{\alpha}

\def\dd{\delta}
\def\DD{\Delta}
\def\ee{\epsilon}

\def\gg{\gamma}
\def\GG{\Gamma}

\def\LL{\Lambda}

\def\rr{\rho}

\def\SS{\Sigma}

\begin{document}
\begin{titlepage}
\begin{flushright}
Bologna preprint \\
DFUB-94-03\\
February 1994\\
hep-ph/xxyyxxx
\end{flushright}
\vspace*{0.5cm}
\begin{center}
{\bf
\begin{Large}
{\bf 
A NEW TOOL FOR THE LAMB-SHIFT CALCULATION\\}
\end{Large}
}
\vspace*{1.5cm}
         {{\large Marco Cavicchi}\footnote{
 E-mail: cavicchi@bologna.infn.it ~/ ~ cavicchi@nbivax.nbi.dk },
\\
          {\large Antonio Vairo}\footnote
{E-mail: vairo@bologna.infn.it}}
         \\[.3cm]
          I.N.F.N. - {\it Sez. di Bologna and Dip. di Fisica,
         \\[.1cm]
          Universit\`a di Bologna, Via Irnerio 46, I-40126 Bologna,
          Italy}\\
\end{center}
\vspace*{0.7cm}
\begin{abstract}
{
We solve the Bethe-Salpeter equation for hydrogenic bound states by choosing
an appropriate interaction kernel $K_c$. We want to use our solution to
calculate up to a higher order the hydrogen Lamb-shift, and as a first
application we present up to order $\left(\aa / \pi\right)(\za)^7$ the
contribution of the lowest order self-energy graph, calculated {\it exactly}.
The basic formalism is a natural extension to the hydrogenic bound states of the
one previously presented by R. Barbieri and E. Remiddi and used in the case of
positronium.
}
\end{abstract}
\vfill
\end{titlepage}

\setcounter{footnote}{0}
\def\ut{{\tilde u}}
\def\zt{{\tilde z}}
\def\dz{{\sqrt{2}z}}

\def\uij{U_{ij}}
\def\ucij{U^\dagger_{ij}}
\def\uji{U_{ji}}
\def\ucji{U^\dagger_{ji}}
\def\dag{\dagger}
\def\zpm{z_{\pm}}
\def\zp{z_+}
\def\zm{z_-}
\def\ddt{{\dd T}}
\def\mucr{\mu_{cr}}

\newcommand{\mat}[4]{\left( 
                     \begin{array}{cc}
                     {#1} & {#2} \\
                     {#3} & {#4} 
                     \end{array}
                     \right)
                    }

\newcommand{\ft}[3]{ {d^{#1}{#2}\over (2\pi)^{#1}} ~ e^{i {#2}\cdot{#3}} }
\newcommand{\ftt}[2]{{d^{#1}{#2}\over (2\pi)^{#1}}}

\def\rhm{\rho_-}
\def\rhp{\rho_+}
\def\sgm{\sigma_-}
\def\sgp{\sigma_+}

\def\rd{\sqrt{2}}
\def\usrd{{1\over\sqrt{2}}}
\def\dxy{\delta^2(x-y)}
\def\dij{\delta^{ij}}
\def\dsi{\partial_{x^-}}
\def\dta{\partial_{x^+}}

\newcommand\modu[1]{|{#1}|}
\newcommand\fai[4]{{#1}^{{#2}}_{{#3}} ({#4}) }

\def\ggv{{\vec \gg}}
\def\pv{{\vec p}}
\def\qv{{\vec q}}
\def\kv{{\vec k}}

\sect{Introduction}

Bound state systems like positronium and muonium are a good test of QED.
Hydrogenic atoms are not completely reducible to a QED problem 
because the structure of the proton. In particular the finite size 
of the proton gives rise to a lower limit of the theoretical calculus 
precision (for a good review see~\cite{Ki} and more recently~\cite{DFT}).

However, there are two kind of considerations which make hydrogenic 
atoms still interesting from a QED point of view. First, the requested 
precision is not yet achieved in many theoretical predictions, e.g. 
the hydrogen Lamb-shift requires a theoretical precision 
of about 1 kHz (the present status of the calculation of the 
hydrogen Lamb-shift can be found in~\cite{DGE}), and new but incomplete
contributions have been recently calculated (e.g.~\cite{EG},~\cite{EKS}).
Second, the fully detailed treatment of a relatively simpler problem 
as the hydrogenic atom (which is reducible in QED to a one particle 
problem in an external potential) can give a hint for the treatment of 
more complicated system as the positronium. For this purpose it is necessary 
to use the same formalism to describe the positronium and the hydrogenic 
atoms. 

In sect.2 we extend to the hydrogenic atoms the formalism
which was proposed by R. Barbieri and E. Remiddi~\cite{BR} for the 
positronium (which we call BR formalism). Following~\cite{BR} we propose 
as interaction kernel a sort of {\it relativistic dressed} 
Coulomb interaction, so that the Bethe-Salpeter equation, which for 
bare Coulomb interaction is the Dirac-Coulomb equation, is analytically 
solvable in closed form. Then we write a perturbative expansion for the energy
levels which immediately reproduces the Dirac levels.
In sect.3, as a first application of our formalism, we calculate 
analytically up to order $\aa / \pi (\za)^7$ for the levels $n = 1,2$ 
the first contribution to the self-energy together with other graphs 
necessary to cancel spurious terms, which typically arise in this 
sort of calculation.
An appendix is devoted to review and discuss the method proposed by 
J. Sapirstein ~\cite{Sa1} to treat perturbatively the Dirac propagator. 
\\

\sect{BR formalism in the context of the hydrogenic atoms}

In QED hydrogenic atoms are well approximated by an electron moving 
in an external field $V_c(r) = -\za/r$, that represents
his Coulomb interaction with the nucleus.
The Green's function $G(W;\pv,\qv)$ of this electron contains all the necessary
informations about the system, and it can be written
in perturbation theory as the sum of all the 
Feynman's graphs with an incoming and an outcoming electron leg
(see fig.1; $W$ is the energy and $\pv, \qv$ are the 
incoming and outcoming spatial components of the electron's momentum;
the ball includes all radiative corrections, the external lines are Coulomb
interaction vertices).
\p
\p
\p
\\
\makebox[4truecm]{\phantom b}
\epsfxsize=7truecm
\epsffile{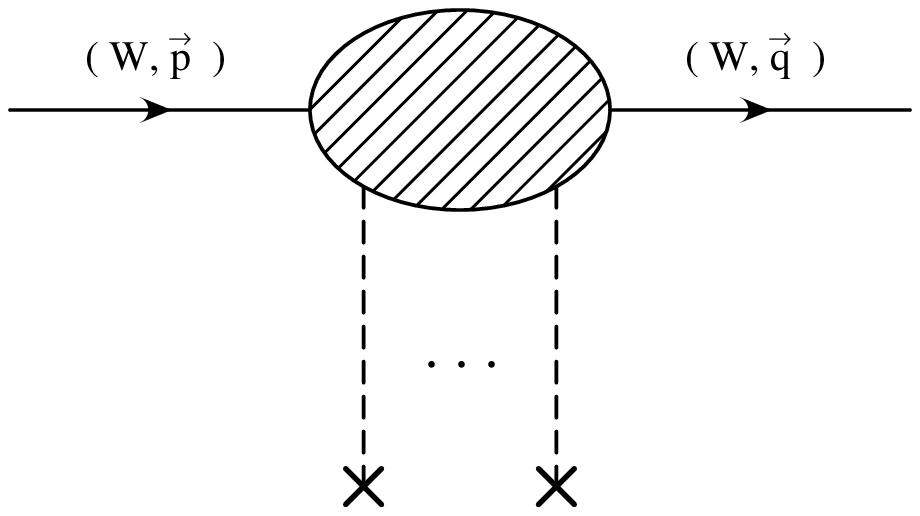}
\\
\p
\p
\centerline{Fig.1 \it{The Green's function} $G(W;{\vec p},\qv)$ \it{of an 
electron in external field}.}
\vskip 1truecm
It is well known that in bound state theory the Green function $G$ has simple 
poles in the energy $W$. For hydrogenic atoms these poles are grouped round 
the Balmer levels $-m(\za)^2/2n^2$ and are labeled by the quantum numbers $n$, 
$l$ (the angular momentum which is an `` almost good quantum number" in the 
sense of ~\cite{BS2}) and $j$ (the total momentum). 
For every pole $W_{nlj}$ we can write:
\eq 
G(W;\pv,\qv) = {{R_{nlj}(\pv,\qv)}\over{W-W_{nlj}}} + 
{\hat G}_{nlj}(W;\pv,\qv) ,
\lbl{g}
\en
where $R_{nlj}$ is the residuum and ${\hat G}_{nlj}$ is the regular part 
of $G$ at $W=W_{nlj}$. 

The standard way to obtain a perturbative expansion of $W_{nlj}$ 
starts from the Bethe-Salpeter equation~\cite{BS}:
\eqa 
G(W;\pv,\qv)&=& G_0(W;\pv){\Bigg[} (2\pi)^3\dd^{(3)}(\pv-\qv)
\nonumber\\
&+& \int\ftt{3}{k}K(W;\pv,\kv)G(W;\kv,\qv)\Bigg] ,
\lbl{bs}
\ena
where $G_0(W;\pv) = i ( \gg^0W-\pv \cdot {\ggv}-m+i\ee)^{-1}$ 
is the electron free propagator and $K(W;\pv,\qv)$ is the interaction kernel
(i.e. the sum of all 1-particle irreducible graphs with external fermionic
legs removed). Then we write:
\eq
K(W;\pv,\qv) \equiv K_c(W;\pv,\qv) + \dd K(W;\pv,\qv) ,
\lbl{dk}
\en
which is, rigorously speaking, a definition of $\dd K$, once $K_c$ is 
suitably chosen. To choose $K_c$, we ask that it satisfies the two conditions:
\\
$(i)$ \  in the non relativistic limit 
$K_c \to -i\gg_0 V_c$; 
\\
$(ii)$ the Bethe-Salpeter equation for $K_c$:
\eqa 
~~~~~~~~~G_c(W;\pv,\qv) &=& G_0(W;\pv){\Bigg[} (2\pi)^3\dd^{(3)}(\pv-\qv)
\nonumber\\
&+& \int\ftt{3}{k}K_c(W;\pv,\kv)G_c(W;\kv,\qv)\Bigg] ,
\lbl{bsc}
\ena
can be explicitly solved in closed form.

From condition $(i)$ it follows that $G_c$ can be considered
as a zeroth order approximation of $G$. Therefore $G_c$ has singularities in 
$W=W^c_n$ 
\footnote{According to general
expectations the exact problem has less degeneracy than the unperturbed
one.}
and $W^c_n$ $\approx$ $m - m(\za)^2/2n^2$:
\eq 
G_c(W;\pv,\qv) = {{\sum_{lj} R^c_{nlj}(\pv,\qv)}\over{W-W^c_n}} + 
{\hat G}^c_n(W;\pv,\qv) ,
\lbl{gc}
\en
where $\sum_{lj} R^c_{nlj}$ is the residuum and ${\hat G}^c_n$ is the 
regular part of $G_c$ at $W=W^c_n$. From condition $(ii)$ it follows 
that $W^c_n, R^c_{nlj}$ and ${\hat G}^c_n$ are explicitly known. 
Using the quantities defined above, first we can write the Bethe-Salpeter
equation  (\ref{bs}) in terms of $G_c$ and $\dd K$ as:
\eqa 
G(W;\pv,\qv) &=& G_c(W;\pv){\Bigg[} (2\pi)^3\dd^{(3)}(\pv-\qv)
\nonumber\\
&+& \int\ftt{3}{k}\dd K(W;\pv,\kv)G(W;\kv,\qv)\Bigg] ,
\lbl{bs2}
\ena
and then the perturbative expansion of the energy levels:
\eqa
W_{nlj} = W^c_n 
&+& {1\over D}Tr\left\{\dd K(W^c_n)R^c_{nlj}\right\}  
\nonumber\\ 
&+& {1\over D^2}Tr\left\{\dd K(W^c_n) {\hat G}^c_n(W^c_n) \dd K(W^c_n)
R^c_{nlj}\right\} 
\lbl{energy}\\ 
&+& {1\over D^2}Tr\left\{\dd K(W^c_n) R^c_{nlj}\right\}
Tr\left\{\pd{W}\dd K(W^c_n) R^c_{nlj}\right\} + ... \,\, , 
\nonumber
\ena
where we have omitted for simplicity the explicit indication of the momenta, 
and $D$ is defined as the following trace \footnote{
From (\ref{bsc}) it follows that $G_c^{-1}  = G_0^{-1}-K_c$ and
then we can write (\ref{deg}) also as:
$$
D =-Tr\left\{i\gg^0 R^c_{nlj}\right\}
        -Tr\left\{\pd{W} K_c(W^c_n) R^c_{nlj}\right\} .
$$
}:
\eq
D \equiv Tr\left\{\pd{W} G_c^{-1}(W^c_n) R^c_{nlj}\right\} .
\lbl{deg}
\en
Each term of (\ref{energy}) is a series in $(\za)$ with fixed $(\aa / \pi)$.
We observe that the expansion (\ref{energy}) is ``perturbative" 
in the sense of increasing orders of $\dd K(W^c_n)$. 
For consistency the explicit calculation must exhibit that 
to an increasing order in $\dd K(W^c_n)$
it corresponds an increasing leading order in $\za$. 
In~\cite{RS} the reader can find a more detailed discussion.
\\
In~\cite{CL},~\cite{BR} (see also~\cite{BuR1}) it was found a kernel 
$K_c$ satisfying conditions $(i)$ and $(ii)$ for positronium.
In what follows we choose for $K_c$ \footnote{
An other choice can be $K_c = K_D \equiv -i\gg^0 {\tilde V}_c$. 
As a consequence of 
this choice $G_c = g_D$ (where $g_D$ is the Dirac-Coulomb propagator and will
be defined later). $g_D$ is known in analytical closed form and its poles
and residuum at the poles also (see ~\cite{MOHR} and for a detailed study
~\cite{CRATER}). But analytical computation with these quantities are more
complicate than in our formalism because the unpractical structure of $g_D$.
}:
\eq
K_c(W;\pv,\qv) = -i \  f(W;p,q) {\tilde V}_c(\pv-\qv) \gg^0
{\LL _+}(\pv){{1+\gg^0}\over 2}{\LL _+}(\qv) ,
\lbl{kc}
\en
where ${\tilde V}_c(\pv) = -4\pi\za/\pv\,^2$ is the Fourier 
transform of the Coulomb potential, and
\eqa
{\LL _{\pm}}(\pv) &=& {{E_p \pm (m-\pv \cdot {\ggv})\gg^0}\over{2E_p}} ,
\nonumber\\
f(W;p,q) &=& 
\left({{16m^2E_pE_q}\over{(E_p+m)(E_p+W)(E_q+m)(E_q+W)}}\right)^{1\over 2} ,
\nonumber\\
E_p &=& \sqrt{{\pv}^{~2}+m^2} .
\nonumber
\ena
In the non relativistic limit $ \pv, \qv \mathrel{\mathop{\longrightarrow}} 0$
it is easily seen that $K_c$ satisfies condition $(i)$.

If we define $H_c(W;\pv,\qv)$ as:
\eqa
&\ &G_c(W;\pv,\qv) \equiv G_0(W;\pv){\Bigg[}(2\pi)^3\dd^{(3)}(\pv-\qv)
\nonumber\\
&\ &-i~f(W;p,q)\gg^0{\LL _+}(\pv){{1+\gg^0}\over 2}{\LL _+}(\qv)
G_0(W;\qv) H_c(W;\pv,\qv) \Bigg] ;
\nonumber\\
\lbl{hc}
\ena
from the comparison of (\ref{hc}) and (\ref{bsc}) we obtain the following 
equation for $H_c$:
\eqa
H_c(W;\pv,\qv) &=& {\tilde V}_c(\pv-\qv) 
\lbl{bsh}\\
&+&\int\ftt{3}{k}
{1 \over { {{W^2-m^2} \over {2m}} - {{\kv^{~2}}\over {2m}} +i\ee}}
{\tilde V}_c(\pv-\kv) H_c(W;\kv,\qv) ;
\nonumber
\ena
the solution of (\ref{bsh}) is known and can be written in the Schwinger 
integral representation~\cite{S}:
\eqa
H_c(W;\pv,\qv) &=& {\tilde V}_c(\pv-\qv) 
+ (\za)^2{1\over{(\pv-\qv)^2}}{{4\pi m }\over{ \sqrt{m^2-W^2} }}
\nonumber\\
&\cdot& 
\int^1_0 d\rr 
{ \rr^{-m\za \over \sqrt{m^2-W^2}} \over
\rr + { (E_p^2-W^2)(Eq^2-W^2)\over 4(m^2-W^2)(\pv-\qv)^2} (1-\rr)^2 } .
\lbl{hcs}
\ena
Inserting (\ref{hcs}) into (\ref{hc}) we obtain:
\eqa
G_c(W;\pv,\qv) &=& (2\pi)^3\dd^{(3)}(\pv-\qv)G_0(W;\pv)
\nonumber\\
&+&i~{f(W;p,q) \over {(E_p-W)(E_q-W)}}
{\LL _+}(\pv){{1+\gg^0}\over 2}{\LL _+}(\qv)\gg^0{\tilde V}_c(\pv-\qv)
\nonumber\\
& \cdot & \left[ 1+\za {m \over \sqrt{m^2-W^2}} \int^1_0 d\rr 
{ \rr^{-m\za \over \sqrt{m^2-W^2}} \over
 \rr + { (E_p^2-W^2)(E_q^2-W^2)
       \over 
        4(m^2-W^2)(\pv-\qv)^2}(1-\rr)^2}
\right] ,
\nonumber\\
\lbl{gcs}
\ena
so that condition $(ii)$ is also fulfilled.
\\
From (\ref{gcs}), 
we see how the Green function $G_c$ has poles at the values of $W$ 
\eq
W^c_n = m\sqrt{1-{{(\za)^2}\over{n^2}}} ,
\lbl{wc}
\en
and isolating the singular part from the regular one at $W=W^c_n$ in 
(\ref{gcs}) we find also the residuum $\sum_{lj} R^c_{nlj}$ at the pole:
\eqa
\sum_{lj}R^c_{nlj}(\pv,\qv) &=& {i \over {W^c_n}}
\left( {{E_p(E_p+W^c_n)} \over {E_p+m}} \right)^{1 \over 2}
\left( {{E_q(E_q+W^c_n)} \over {E_q+m}} \right)^{1 \over 2}
\nonumber\\
&\cdot&{\LL _+}(\pv){{1+\gg^0}\over 2}
\sum_{l=0}^{n-1}R_{nl}(p)R_{nl}(q)
{{2l+1}\over{4\pi}}P_l\left({{\pv\cdot\qv}\over{pq}}\right) 
{\LL _+}(\qv)\gg^0 ,
\nonumber\\
\lbl{res}
\ena
where $P_l(z)$ is the Legendre polynomial of order $l$ and
$R_{nl}$ the radial part of the Schr\"odinger-Coulomb wave functions.
\\
${\hat G}^c_n(W;\pv,\qv)$, the regular part of $G_c$ at $W=W^c_n$, can 
be obtained explicitly subtracting from (\ref{gcs}) the singular 
part of (\ref{gc})  and taking (\ref{res}) into account (see~\cite{BuR2}
for the explicit case of positronium).

To obtain from (\ref{res}) the expression of $R^c_{nlj}$, following 
~\cite{RS}, we write the identity:
\eq
{{2l+1}\over{4\pi}}P_l\left({{\pv\cdot\qv}\over{pq}}\right) = 
\sum_{j=|l-{1\over 2}|}^{j=l+{1\over 2}} D^j_l
\left({\pv \over p},{\qv \over q}\right) , 
\nonumber
\en
where 
\eqa
D^j_l\left({\pv \over p},{\qv \over q}\right) &=& {1\over {4\pi}}
\left(j+{1\over 2}\right)P_l\left({{\pv\cdot\qv}\over{pq}}\right)
\nonumber\\
&+&{{j-l}\over{2\pi}}
\left( {{\pv \cdot {\ggv}}\over p}{{\qv \cdot {\ggv}}\over q}
+ {{\pv \cdot \qv}\over {pq}} \right)
\left.\pd{z}P_l(z)\right|_{\cos z = {{\vec p}\cdot{\vec q}\over pq}}  .
\nonumber
\ena
$R^c_{nlj}$ is then given by:
\eqa
R^c_{nlj}(\pv,\qv) &=& {i \over {W^c_n}}
\left( {{E_p(E_p+W^c_n)} \over {E_p+m}} \right)^{1 \over 2}
\left( {{E_q(E_q+W^c_n)} \over {E_q+m}} \right)^{1 \over 2}
\lbl{resj}\\
&\cdot& R_{nl}(p)R_{nl}(q) 
{\LL _+}(\pv){{1+\gg^0}\over 2}
D^j_l\left({\pv \over p},{\qv \over q}\right)
{\LL _+}(\qv)\gg^0      ,
\nonumber
\ena
which is an eigenfunction of parity and total momentum.
\\
$R^c_{nlj}$ satisfies:
\eq 
R^c_{nlj}(\pv,\qv) = 
G_0(W_n^c;\pv)\int\ftt{3}{k}K_c(W_n^c;\pv,\kv)R^c_{nlj}(\kv,\qv),
\lbl{bsrc}
\en
known as the Bethe-Salpeter equation for the residuum.

The expansion (\ref{energy}) is now completely defined. The first terms of 
the expansion that must be calculated are $D$ and 
$Tr\left\{K_c(W^c_n)R^c_{nlj}\right\}$. These terms are not related to any 
Feynman graph, they only depend on the formalism which we have adopted. 
\\
After explicit calculation we find:
\eq
D = 2j+1 ,
\lbl{deg1}
\en
and 
\eq
\left<K_c\right>_{n} \equiv 
{1\over D}Tr\left\{K_c(W^c_n)R^c_{nlj}\right\} = -{{m^2}\over{W_n^c}}
{{(\za)^2}\over{n^2}} .
\lbl{kclevel}
\en
$D$ is the degeneration of the level $(n,l,j)$ 
\footnote{This result less than surprising is a natural consequence 
of the Bethe-Salpeter formalism.} and $\left<K_c\right>_{n}$ is, 
at the leading order (apart for a factor 2), the Balmer series
(which follows from condition $(i)$ on $K_c$).

The next term of the expansion (\ref{energy}) giving the leading  
corrections to the level is $Tr\left\{K(W^c_n)R^c_{nlj}\right\}$. 
In fig.2 we show $K$ up to one loop radiative corrections.
\p
\p
\p
\\
\makebox[1.5truecm]{\phantom b}
\epsfxsize=12truecm
\epsffile{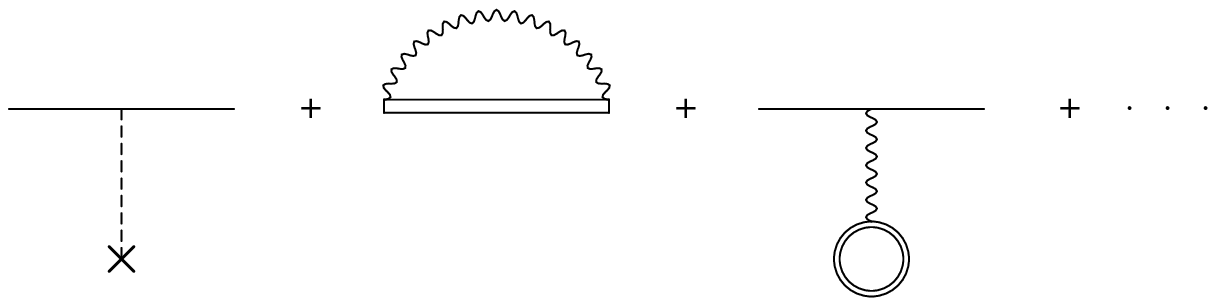}
\\
\p
\p
\centerline{Fig.2 \it{The first terms of the irreducible kernel $K$ 
for hydrogenic atoms}.}
\vskip 1truecm
In fig.2 with the doubled line we have represented the sum of graphs of fig.3.
These graphs can be resummed using the Dirac-Coulomb equation:
\eqa
g_D(W;\pv,\qv) &=& G_0(W;\pv)\Bigg[
(2\pi)^3\dd^{(3)}(\pv-\qv)
\nonumber\\
&+& \int\ftt{3}{k}\left(-i{\tilde V}_c(\pv-\kv)\gg^0\right)g_D(W;\kv,\qv)
\Bigg] .
\lbl{gd}
\ena
This equation is of the kind of the Bethe-Salpeter equation (\ref{bs})
with kernel:
\eq
K_D(\pv,\qv) = -i{\tilde V}_c(\pv-\qv)\gg^0 .
\lbl{kd}
\en
A graphical representation of $K_D$ is given by the first graph of fig.2.
Therefore equation (\ref{gd}) can be treated perturbatively in 
the formalism of equations (\ref{bs})-(\ref{bs2}) with $K_c$ given 
by equation (\ref{kc}) and consequently $\dd K$ $=$ $K_D-K_c$.
An alternative approach to treat perturbatively the Dirac-Coulomb propagator 
$g_D$ was given by J. Sapirstein ~\cite{Sa1} and can be found in appendix.
\vskip 1truecm
\p
\p
\p
\\
\makebox[1truecm]{\phantom b}
\epsfxsize=12truecm
\epsffile{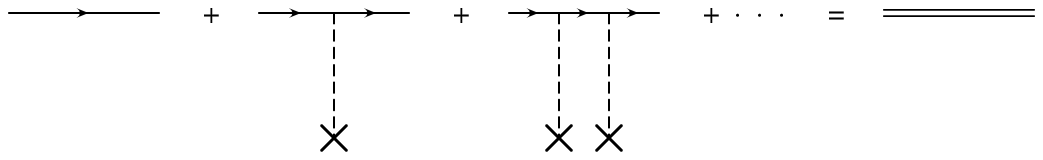}
\p
\p
\centerline{Fig.3 \it{The Dirac-Coulomb propagator} $g_D$.}
\vskip 1truecm

The contribution to the energy levels of $K_D$ originates up to order 
${(\za)^4}$ the well-known Dirac levels $W^D_{nj}$:
\eqa
\left<K_D\right>_{nlj} &\equiv& {1\over{D}}\int\ftt{3}{p}\ftt{3}{q}
Tr\left\{-i{\tilde V}_c(\pv-\qv)\gg^0R^c_{nlj}(\pv,\qv)\right\} 
\nonumber\\
&=&-m{{(\za)^2}\over{n^2}}-m{{(\za)^4}\over{2 n^3}}\left( 
{1\over {j+{1\over 2}}}\right) ,
\lbl{v1}\\
W^D_{nj} &=& W^c_n+\left<K_D\right>_{nlj}-\left<K_c\right>_{n}
\nonumber\\
&=&m-m{{(\za)^2}\over{2 n^2}}-m{{(\za)^4}\over{2 n^3}}\left( 
{1\over {j+{1\over 2}}} -{3\over {4 n}} \right) .
\lbl{wd}
\ena
More interesting are the one loop corrections. In particular
in the next section we discuss the contribution to (\ref{energy}) of 
the second graph of fig.2, the self-energy graph.
\\

\sect{Application to the self-energy contribution}

The first one-loop correction to the hydrogenic atoms' energy levels is given
by the self-energy graph (second graph of fig.2; a recent calculation
of this contribution in an other method can be found in~\cite{Pa}). 

Let us recall that the contribution to the self-energy kernel due to
the first graph of the right hand side of fig.4 could be written as:

\eq
i\SS(p) = 
\int\ftt{4}{k} (-ie\gg^\mu){i\over \psl-\ksl-m+i\ee}(-ie\gg^\nu)
{-ig_{\mu\nu}\over k^2+i\ee} .
\lbl{sig}
\en
\p
\p
\p
\\
\makebox[1truecm]{\phantom b}
\epsfxsize=13truecm
\epsffile{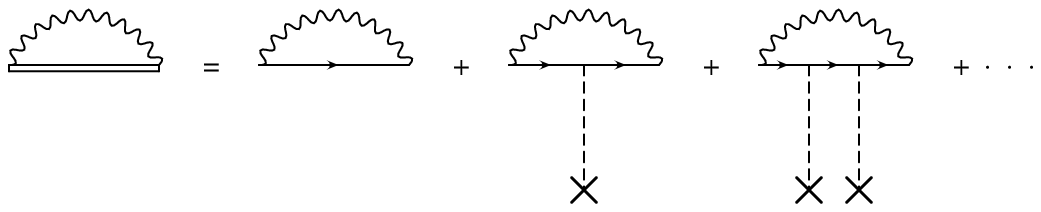}
\p
\p
\centerline{Fig.4 \it{The} $(Z\alpha)$ \it{expansion for the self-energy 
graph}.}
\vskip 1truecm
Expression (\ref{sig}) is obviously divergent.
Choosing the Pauli-Villars regularization one has:
\eqa
i~\SS_{reg}(p;\LL) &=& \int\ftt{4}{k} 
(-ie\gg^\mu){i\over \psl-\ksl-m+i\ee}(-ie\gg^\nu)
\nonumber\\ 
&\cdot& {-ig_{\mu\nu}\over k^2+i\ee} {-\LL^2 \over k^2-\LL^2+i\ee} 
\nonumber\\ 
& =& i~\api[(-\psl+m)A(p^2;\LL) + m~B(p^2;\LL)] ,
\lbl{sig2}
\ena
where
\eqa
A(p^2;\LL) 
       &=&{1\over 4} (1-m^2/p^2) (1+m^2/p^2) 
         \log(1-p^2/m^2)- {1\over 4}\left({m^2\over p^2}-1\right)
\nonumber\\
       &-&{5\over 8} -{1\over4}\llog ,
\nonumber\\
B(p^2;\LL) 
       &=&{1\over 4} (1-m^2/p^2) (3-m^2/p^2) 
         \log(1-p^2/m^2) + {1\over 4}\left({m^2\over p^2} -1\right)
\nonumber\\
       &-&{3\over 8} - {3\over 4}\llog .
\lbl{se}
\ena
The mass-shell renormalization prescription is imposed by defining
the mass renormalization counterterm:
\eq
i~\dd m \equiv -i\SS_{reg}(p;\LL) \biggr\vert_{\psl=m}
\en
so that~\cite{MOHR}
\eq
i~\dd m = i{\aa\over\pi} m \left[ {3\over 8} +  {3 \over 4}\llog\right]
\en  
The relevant kernel for our purpose is the regularized and mass-subtracted 
graph:
\eq
K_{s.e.}(p;\LL) = i\left(\SS_{reg}(p;\LL)+\dd m \right),
\lbl{sig3}
\en
which is still u.v. divergent (wave-function renormalization has not been 
carried out). If we rearrange the terms to isolate the $\LL$ dependent 
part $K^{div}_{s.e.}$ from the rest ${\hat K}_{s.e.}$, we will obtain:
\eq
K_{s.e.}(p;\LL) =  K^{div}_{s.e.}(p;\LL) + {\hat K}_{s.e.}(p) 
\lbl{sig5}
\en
where
\eqa
K^{div}_{s.e.}(p;\LL) &=&i\api(-\psl+m)\left(-{1\over 4}\llog\right)
\nonumber\\ 
&=& \api G^{-1}_0(p)\left(-{1\over 4}\llog\right) ,
\lbl{selfdiv}
\ena
and
\eqa
{\hat K}_{s.e.}(p;\LL)  &=&i~\api[(-\psl+m){\hat A}(p^2) + m~{\hat B}(p^2)] ,
\nonumber\\ 
{\hat A}(p^2) 
      &=&{1\over 4} \left(1-{m^2\over p^2}\right) \left(1+{m^2\over p^2}\right)
         \log(1-p^2/m^2)
\nonumber\\
&-& {1\over 4}\left({m^2\over p^2}-1\right)-{5\over 8},
\nonumber\\
{\hat B}(p^2) 
      &=&{1\over 4} \left(1-{m^2\over p^2}\right) \left(3-{m^2\over p^2}\right)
         \log(1-p^2/m^2) 
\nonumber\\
&+& {1\over 4}\left({m^2\over p^2} -1\right) .
\lbl{sig4}
\ena
The contribution of (\ref{sig5}) to the energy levels according to 
(\ref{energy}) is given by:
\eqa
\left<K_{s.e.}\right>_{nlj} &=& \left<K^{div}_{s.e.}\right>_{nlj} + \left<{\hat K}_{s.e.}\right>_{nlj} 
\nonumber\\
&\equiv& {1\over D}Tr\left\{ K^{div}_{s.e.}(W^c_n) R^c_{nlj} \right\}
+ {1\over D}Tr\left\{ {\hat K}_{s.e.}(W^c_n) R^c_{nlj} \right\} .
\nonumber\\
\ena
Because $R^c_{nlj}$ satisfies the Bethe-Salpeter equation (\ref{bsrc})
and taking in account equation (\ref{selfdiv}) we can easily 
calculate $\left<K^{div}_{s.e.}\right>_{nlj}$:
\eqa
\left<K^{div}_{s.e.}\right>_{nlj} &=&
{1\over D}Tr\left\{ K^{div}_{s.e.} R^c_{nlj} \right\} = 
{1\over D}Tr\left\{ K^{div}_{s.e.} G_0 K_c R^c_{nlj} \right\}
\nonumber\\
&=& \api\left(-{1\over 4}\llog \right){1\over D}
Tr\left\{K_c R^c_{nlj} \right\} 
\nonumber\\
&=& \api\left(-{1\over 4}\llog \right) \left<K_c\right>_{n}.
\lbl{sigdiv}
\ena
The explicit value of $\left<K_c\right>_{n}$ has been given in (\ref{kclevel}).

We don't need to perform a wave-function renormalization: 
if we adopt a gauge invariant regularization scheme
the wave-function renormalization counterterm is exactly compensated
by the vertex renormalization counterterm, hence one obtains
the correct physical result without taking in account these counterterms
\footnote{Moreover in this manner one avoids the problem 
of the spurious infrared divergences which may arise from mass-shell
wave-function renormalization.}.
The $\LL$ dependent term (\ref{sigdiv}) must also cancel
if we evaluate the vertex contribution to the energy
expansion (\ref{energy}) in the same gauge 
invariant Pauli-Villars regularization scheme adopted
for the self-energy and without subtracting the renormalization
counterterm (see for a detailed discussion in the 
positronium context \cite{BuR1}).

We have therefore to consider the second graph of the right hand side 
of fig.4. The kernel due to this graph is:
\eq
K_{ver} = -i{\tilde V}_c \GG^0_{reg} ,
\lbl{kver}
\en
where the Pauli-Villars regularized $\GG^\mu_{reg}$, 
\eqa
\GG^\mu_{reg}(p,q;\LL) &=& \int\ftt{4}{k} (-ie\gg^\rho)
{i\over \psl-\ksl-m+i\ee} \gg^\mu {i\over \qsl-\ksl-m+i\ee}(-ie\gg^\sigma)
\nonumber\\
&\cdot& {-ig_{\rho\sigma}\over k^2+i\ee} {-\LL^2 \over k^2-\LL^2+i\ee},
\ena
satisfies the Ward identity:
\eq
(p-q)_\mu\GG^\mu_{reg}(p,q;\LL) = \SS_{reg}(p;\LL)-\SS_{reg}(q;\LL) .
\lbl{ward}
\en
\\
From (\ref{ward}),(\ref{sig2}) and (\ref{se}), differentiating 
with respect to $p_\nu$ and rearranging terms, it follows that:
\eqa
\GG^\nu_{reg} (p,q;\LL) 
&=&  \pd{p_\nu}\SS_{reg}(p;\LL) 
-(p-q)_\mu\pd{p_\nu}\GG^\mu_{reg}(p,q;\LL)
\nonumber\\
&=& -\api \gg^\nu \left(-{1\over 4}\llog\right) + {\hat \GG^\nu}(p,q) , 
\nonumber\\
\lbl{vtx}
\ena
where ${\hat \GG^\nu}$ is u.v. finite ($\LL$ independent).
\\
The contribution of (\ref{kver}) to the energy levels is also given by:
\eqa
\left<K_{ver}\right>_{nlj} &\equiv&
{1\over D}Tr\left\{ -i{\tilde V}_c\GG^0_{reg}(W_n^c) R^c_{nlj} 
\right\}
\lbl{vtx2}\\ 
&=& -\api\left( -{1\over 4}\llog\right)\left<K_D\right>_{nlj} 
+ \left<{\hat K}_{ver}\right>_{nlj} ,
\nonumber
\ena
where we have defined ${\hat K}_{ver}$ $\equiv$ $-i{\tilde V}_c{\hat\GG}^0$.
The explicit value of $\left<K_D\right>_{nlj}$ has been given in (\ref{v1}).

Comparing (\ref{sigdiv}) with (\ref{vtx2}) we note that only the 
$\aa/ \pi(\za)^2$$\log(\LL^2/m^2)$  terms, corresponding to the 
leading order in $\za$, cancel. Then for a complete cancellation of 
the divergent terms it is necessary to sum other divergent terms arising 
from expansion (\ref{energy}) \footnote{
We note that this problem doesn't occur if we use the kernel 
$K_c = K_D$ in this case the divergent part of (\ref{vtx2}) completely 
cancel (\ref{sigdiv}).}.
To obtain for instance the cancellation up to order 
$\aa/\pi(\za)^4$$\log(\LL^2/m^2)$ the terms which must be considered arise 
from the last contribution explicitly written in expansion (\ref{energy}):
\par
$$
\left<\dd K\right>_{nlj}\left<\pd{W}\dd K\right>_{nlj} \equiv 
{1\over{D^2}}Tr\left\{\dd K(W^c_n) R^c_{nlj}\right\}
Tr\left\{\pd{W}\dd K(W^c_n) R^c_{nlj}\right\} .
$$
Because (taking in account (\ref{selfdiv}),(\ref{vtx}) and note 2):
\par
\eqa
\left.\left<\pd{W}\dd K\right>_{nlj}
\right|_{{one~loop}\,\llog\,{terms}} &=& 
\left<\pd{W}K^{div}_{s.e.}\right>_{nlj} ~~~~~~~~
\lbl{derW}\\
=\api \left( -{1\over 4}\llog \right)&\cdot& \left< -i\gg^0 \right>_{nlj}
\nonumber\\
=\api \left( -{1\over 4}\llog \right) &\cdot& 
  \left( 1 + \left<  \pd{W} K_c  \right>_{nlj} \right)
\nonumber\\
=\api \left( -{1\over 4}\llog \right) &\cdot& 
  \left( 1 + O \left( (\za)^2 \right) \right) ;
\nonumber
\ena
one has:
$$
\left.\left<K_D-K_c\right>_{nlj}\left<\pd{W}\dd K\right>_{nlj} 
\right|_{{one~loop}\,\llog\,{terms}} = 
$$
\eq
\left( \left<K_D\right>_{nlj}-\left<K_c\right>_{n}\right)
           \api \left( -{1\over 4}\llog \right) 
+ O\left(\api(\za)^6\llog\right) .
\lbl{WderW}
\en
\\
Summing now (\ref{sigdiv}), (\ref{vtx2}) and
(\ref{WderW}) up to order $\aa/\pi(\za)^4$ all $\log(\LL^2/m^2)$ dependent 
terms cancel.

The contribution of $\left<{\hat K}_{s.e.}\right>_{nlj}$ has been 
evaluated on the levels $n = 1,2$ {\it exactly}, as a first step we show 
only the leading order terms:
\eq
\left<{\hat K}_{s.e.}\right>_{1s} = 
\api(\za)^2\left[-{3\over 8} + 2\log(\za)+ 2\log(2)\right],
\lbl{1s}
\en
\eq
\left<{\hat K}_{s.e.}\right>_{2s} =
\api(\za)^2\left[-{1\over 32} + {1\over 2}\log(\za)\right],
\lbl{2s}
\en
\eq
\left<{\hat K}_{s.e.}\right>_{2p} =
\api(\za)^2\left[-{11\over 96} + {1\over 2}\log(\za) \right].
\lbl{2p}
\en
Here and in the following the indication of just the $nl$ levels means
there isn't contribution to the splitting in the $j$ levels.

The $\aa/\pi(\za)^2$ and $\aa/\pi(\za)^2\log(\za)$
terms in (\ref{1s})-(\ref{2p}) are expect to vanish. 
This occur because the Dirac levels (\ref{wd}) of order $(\za)^4$ 
are completely given by the graphs discussed in the previous section.
This is a well known feature of the Feynman gauge 
(see e.g~\cite{love}); it decreases the speed of convergence of 
the perturbative series by generating spurious terms of low 
order that only at the end of {\it all} 
calculation (up to requested order in $\za$) must cancel each other.

Up to order $\aa /\pi(\za)^2$ the contributions to the energy levels 
$n = 1,2$ of $\left<{\hat K}_{ver}\right>_{nlj}$ are:
\eq
\left<{\hat K}_{ver}\right>_{1s} = \api(\za)^2
\left[{39\over 8} - 2\log(\za)- 10\log(2) \right] ,
\lbl{v1s}
\en
\eq
\left<{\hat K}_{ver}\right>_{2s} = \api(\za)^2
\left[{1003\over 288} - {1\over 2}\log(\za)- {16\over 3}\log(2) \right] ,
\lbl{v2s}
\en
\eq
\left<{\hat K}_{ver}\right>_{2p} = \api(\za)^2
\left[{2009\over 864} - {1\over 2}\log(\za)- {32\over 9}\log(2) \right] .
\lbl{v2p}
\en

These contribution only cancel the $\aa/\pi(\za)^2\log(\za)$ 
spurious terms of (\ref{1s})-(\ref{2p}). To obtain the full cancellation
of all $\aa/\pi(\za)^2$ terms which are in (\ref{1s})-(\ref{2p}) it is
necessary to calculate the leading order of the remaining graphs
of fig.4:
\eqa
\left<K_{\it ladder}\right>_{nlj}  &\equiv& {1\over D}\sum_{j=1}^{\infty} 
\int\ftt{4}{k} {-ig_{\mu\nu}\over k^2+i\ee} \int\ftt{3}{p} \int\ftt{3}{q} 
\nonumber\\
&\ &\int\ftt{3}{x_1} \int\ftt{3}{x_2}... \int\ftt{3}{x_j} 
Tr\left\{ (-ie\gg^\mu) \right.
\nonumber\\
&\ &G_0(W_n^c-k^0,{\vec q}-{\vec k}) K_D(\vec x_1,{\vec q})
G_0(W_n^c-k^0,\vec x_1-{\vec k}) ... 
\nonumber\\
&\ & G_0(W_n^c-k^0,\vec x_j-{\vec k})  K_D(\vec x_j,{\vec p})
G_0(W_n^c-k^0,{\vec p}-{\vec k}) 
\nonumber\\
&\ &~~~~~~~~~~~~~~~~~~~~~~~~~~~~~~~~~~~~~~~~~
\left. (-ie\gg^\nu) R^c_{nlj}(\pv,\qv) \right\}
\nonumber\\
&\approx& {1\over 2n}\api(\za)^2
\int^\infty_0dk\int {d^3p\over (2 \pi)^3} {d^3q\over (2 \pi)^3}
R_{nl}(p)R_{nl}(q) 
\nonumber\\
&\ &P_l\left({\pv\cdot\qv\over pq}\right){1\over(\pv-\qv)^2}
{{p^2+q^2+2}\over (k+p^2+1)(k+q^2+1)}
\lbl{ladder}
\ena
From the explicit calculation of (\ref{ladder}) on the levels 
$n=1$ and $n=2$ it follows:
\eq
\left<K_{\it ladder}\right>_{1s} = 
\api(\za)^2\left[-{9\over 2} + 8\log(2) \right] ,
\lbl{g1s}
\en
\eq
\left<K_{\it ladder}\right>_{2s} = \api(\za)^2
\left[-{497\over 144} + {16 \over 3}\log(2) \right] ,
\lbl{g2s}
\en
\eq
\left<K_{\it ladder}\right>_{2p} = \api(\za)^2
\left[-{955\over 432} + {32\over 9}\log(2) \right] .
\lbl{g2p}
\en
These contributions eliminate, as we have announced, the remaining 
\\
$\aa/ \pi(\za)^2$ terms from (\ref{1s}), (\ref{2s}), (\ref{2p}) and
(\ref{v1s}), (\ref{v2s}), (\ref{v2p}).
\\

At the end of this section we present the complete (but for short up to order 
$\left(\aa / \pi \right)(\za)^7$) contribution of 
$\left<{\hat K}_{s.e.}\right>_{nlj}$ without the terms cancelled by the 
leading contribution of $\left<{\hat K}_{ver}\right>_{nlj}$ and 
$\left<K_{ladder}\right>_{nlj}$:
\\
\eqa
\DD E_{1s} &\equiv& \Big<{\hat K}_{s.e.}+
\left.\left({\hat K}_{ver}+K_{ladder} \right) 
\right|_{leading}\Big>_{1s} =
\nonumber\\
&\ &\api(\za)^4\left[ {37\over 16} + 7\log(\za)+ 7\log(2)\right]
\nonumber\\
&+&\api(\za)^5\left[{20\over {9\pi}}-6\pi-{1\over \pi}I-{14\over 3\pi}\log(2)
\right] 
\nonumber\\
&+&\api(\za)^6\left[{983\over 64}
-{45\over 4}\log(\za)-{45\over 4}\log(2)\right]
\nonumber\\
&+&\api(\za)^7\left[{1237\over 225\pi} - {5\over 4\pi}I 
- {463\over 30 \pi}\log(2) 
\right] ;
\nonumber\\
\lbl{1scom}
\ena
\eqa
\DD E_{2s} &\equiv& \Big<{\hat K}_{s.e.}+
\left.\left({\hat K}_{ver}+K_{ladder} \right) 
\right|_{leading}\Big>_{2s} =
\nonumber\\
&\ &\api(\za)^4\left[ {191\over 256} + {13\over 16}\log(\za)\right]
\nonumber\\
&+&\api(\za)^5\left[{5\over 18\pi}-{3\pi\over 4}-{1\over 8\pi} I 
  - {7\over 12\pi}\log(2)\right]
\nonumber\\
&+&\api(\za)^6\left[{2615\over 12288}-{509\over 256}\log(\za) \right]
\nonumber\\
&+&\api(\za)^7\left[{133\over 7200\pi}+{3\pi\over 4}-{53\over 128\pi}I 
-{607\over 960\pi}\log(2) \right] ;
\nonumber\\
\lbl{2scom}
\ena
\eqa
\DD E_{2p} &\equiv& \Big<{\hat K}_{s.e.}+
\left.\left({\hat K}_{ver}+K_{ladder} \right) 
\right|_{leading}\Big>_{2p} =
\nonumber\\
&\ &\api(\za)^4\left[ -{43\over 768} + {5\over 16}\log(\za)\right] 
\nonumber\\
&+&\api(\za)^6\left[{9181\over 36864} - {329\over 768}\log(\za)\right]
\nonumber\\
&+&\api(\za)^7\left[
{23\over 450\pi} -{\pi \over 4} +{1\over 8\pi}I 
+ {1\over 20\pi}\log(2) \right] ;
\nonumber\\
\lbl{2pcom}
\ena
where we have defined \footnote{
The presence of the dilogarithms in (\ref{i1}) is not surprising, 
being a standard feature of this kind of calculations.
For example, such terms can have their origin from integrals of the kind 
$$
f(W)=-\int_0^1 {dy\over y} \log(y^2+2Wy+1) ,
$$
obtained after the change of variables $p+E_p=y$ and after some 
rationalizations. It is easy to see how 
$$
{df(W)\over dW} = -{1\over \sqrt{1-W^2}}\atan{\sqrt{1-W^2}\over W},
$$
$$
f(1) = 2\Li_2(-1) = -{\pi^2\over 6},
$$
so that 
$$
f(W) = -{\pi^2\over 6} + {1\over 2}\atan^2{\sqrt{1-W^2}\over W}.
$$
Other terms come from similar integrals giving at the end, after an
expansion in $\za$, the results (\ref{1s})-(\ref{2p}).}:
\eq
I = {\Li_2(3-2\sq2)+2\Li_2(1) + \log^2(1+\sq2)\over \sq2}.
\lbl{i1}
\en
\\

\sect{Conclusions}

We have shown how it is possible to perform a perturbative expansion of
the hydrogenic bound state two-points Green function by choosing an appropriate 
zeroth-order kernel. We have written the corresponding zeroth-order solution 
and the perturbative expansion for the bound state energy levels.
Our approach is very transparent and unambiguous in the sense
that in this way one knows exactly what he is discarding and what he is 
keeping, and each approximation is just referred to neglecting some Feynman
graph.
Then we have expanded perturbatively the self-energy, and we have 
calculated the exact contribution of the first graph in the expansion, 
here presented up to order $\left(\aa / \pi \right)(\za)^7$.
The way is still long and hard: the next terms, which contribute to the 
self-energy are the one-Coulomb exchange diagram and the sum 
(from two up to infinity) of {\it all} many-Coulomb exchanges. 
While for the one-Coulomb exchange one may think to proceed in the
calculation by brute force, for the sum of the many-Coulomb exchange graphs 
there are the difficulties to treat the Schwinger integral
which compare in the Dirac-Coulomb propagator's expansion. 
These difficulties are essentially the same which are present in the other 
bound-state problems in QED. Therefore, the solution of these difficulties 
is not only important in order to have progress in the hydrogenic atoms energy 
levels calculation but it will make possible, in particular, an {\it ex novo } 
calculation in our formalism of the positronium energy levels.
Along this direction we are going proceed further on.
\\
\\

{\large \bf Acknowledgments}
\\

We thank Ettore Remiddi, who stimulated our interest on this problem,
for many helpful discussions and for carefully reading this paper.
\\

\appendix
\sect{Appendix}

All graphs of fig.3 give contributions to the energy level of the same 
order in $\za$, so that such an expansion of the Dirac-Coulomb propagator 
$g_D$ is wrong from the perturbative point of view. In section 2 we have 
discussed how to treat perturbatively in our formalism the Dirac-Coulomb
propagator. 
In this appendix we review the alternative method proposed by J. Sapirstein
~\cite{Sa1} (see~\cite{Sa2} for an application). 
This method is more artificial than ours which follows as a natural consequence 
from the formalism. Therefore one expects to have some additional
analytical problems to implement the following expansion for $g_D$ into
the energy expansion (\ref{energy}). 
\\
The idea is to isolate from $g_D$ a scalar part $s_D$ and then to write 
a perturbative expansion for $s_D$. We define $s_D$ as:
\eqa
g_D(W;\pv,\qv) &\equiv& (\gg^0 W-\pv\cdot{\ggv}+m)s_D(W;\pv,\qv)
\nonumber\\ 
&+& {{\za\gg^0}\over{2\pi^2}}\int d^3k{1\over{(\pv-\kv)^2}}s_D(W;\kv,\qv)
\nonumber\\ 
&=& (iG_0^{-1}-{\tilde V}_c\gg^0)s_D .
\lbl{sd}
\ena
This means also that $s_D = -ig_D^2$.
\\
In the following, in order to make the notation short, we will write all
formulas as in (\ref{sd}), i.e. without the explicit
indication of the momenta and their integration.
Substituting (\ref{sd}) into the Dirac-Coulomb equation (\ref{gd}) we 
obtain the equation for $s_D$:
\eqa
s_D &=& g_0(1+k\ s_D) ,
\lbl{bsd}\\
k&=&k_c+\dd k ,
\lbl{kkd}
\ena
where:
\eqa
g_0(W;\pv,\qv)&\equiv&-{i\over{E_p^2-W^2}}
(2\pi)^3\dd^{(3)}(\pv-\qv) ,
\nonumber\\
k_c(W;\pv,\qv)&\equiv&8\pi i\ \za {W\over{(\pv-\qv)^2}} ,
\nonumber\\
\dd k(\pv,\qv)&\equiv&2\pi^2 i (\za)^2{1\over{|\pv-\qv|}} + 
4\pi i (\za){{\gg^0~{\ggv}\cdot(\pv-\qv)}\over{(\pv-\qv)^2}}
\nonumber\\
&\equiv&\dd k_1 + \dd k_2 .
\nonumber
\ena
We can consider (\ref{bsd}) and (\ref{kkd}) 
as the analogous of (\ref{bs}), (\ref{dk}) respectively.
\\
It is important to remark that in (\ref{kkd}) $\dd k$ is really
a ``perturbation'' of $k_c$ in the sense that all the contributions which arise 
from $\dd k$ are of higher order in $\za$ than the contributions which arise 
from $k_c$. 
Finally we can write as in (\ref{bs2}) the Bethe-Salpeter equation for $s_D$:
\eq
s_D = g_c(1+\dd k\ s_D) ,
\lbl{bsd2}
\en
where $g_c$ is given by (see (\ref{gcs})):
\eqa
g_c(W;\pv,\qv) &=& g_0(W;\pv,\qv) 
-{1\over{E_p^2-W^2}}k_c(W;\pv,\qv){1\over{E_q^2-W^2}}
\nonumber\\
& \cdot & \left[ 1+\za {W \over {\sqrt{m^2-W^2}}} \int^1_0 d\rr 
{{\rr^{-{W\za}\over{\sqrt{m^2-W^2}}}} \over
 {\rr + (1-\rr)^2{ {(E_p^2-W^2)(Eq^2-W^2)}\over 
                 {4(m^2-W^2)(\pv-\qv)^2}}}}\right]
\nonumber\\
&\equiv&g_0+g^c_1+g^c_M .
\lbl{gcs2}
\ena
From (\ref{gcs2}), (\ref{bsd2}) and (\ref{sd})  we obtain a perturbative
expansion for the Dirac-Coulomb propagator $g_D$:
\eqa
g_D &=& (iG_0^{-1}-{\tilde V}_c\gg^{0})\Bigg[g_c + g_c~\dd k~g_c + 
\nonumber\\
&+& g_c~\dd k~g_c~\dd k~(g_c + g_c~\dd k~g_c+...)\Bigg] .
\lbl{gd1}
\ena
If we want to isolate from this expansion the first two terms 
of fig.3, we observe that:
\eq
iG_0^{-1}g_0 = (2\pi)^3\dd^{(3)}(\pv-\qv)G_0 ,
\en
\eq
iG_0^{-1} g^c_1 + iG_0^{-1} g_0~\dd k_1~g_0 
- {\tilde V}_c\gg^0 g_0
= G_0 (-i\gg^0{\tilde V}_c) G_0 ,
\en
and hence we rewrite (\ref{gd1}) as:
\eqa
g_D &=& G_0 + G_0 (-i\gg^0{\tilde V}_c) G_0 + iG_0^{-1} g^c_M
\nonumber\\ 
&+& iG_0^{-1} g_0~\dd k_2~g_0
-{\tilde V}_c\gg^{0}(g^c_1+g^c_M)-{\tilde V}_c\gg^0g_0~\dd k~g_0
\nonumber\\
&+& (iG_0^{-1}-{\tilde V}_c\gg^0)\Bigg[(g^c_1 + g^c_M)~\dd k~g_c 
+ g_c~\dd k~(g^c_1 + g^c_M)
\nonumber\\
&+& g_c~\dd k~g_c~\dd k~(g_c + g_c~\dd k~g_c+...)\Bigg] .
\lbl{gd2}
\ena
We observe that the leading order of the contribution of the remaining 
graphs of fig.3 arises from $iG_0^{-1}g^c_M$.
\\

\end{document}